# Burst intensification by singularity emitting radiation in multi-stream flows

A. S. Pirozhkov[1*], T. Zh. Esirkepov[1*], T. A. Pikuz[2,3], A. Ya. Faenov[3,4], K. Ogura[1], Y. Hayashi[1], H. Kotaki[1], E. N. Ragozin[5,6], D. Neely[7,8], H. Kiriyama[1], J. K. Koga[1], Y. Fukuda[1], A. Sagisaka[1], M. Nishikino[1], T. Imazono[1], N. Hasegawa[1], T. Kawachi[1], P. R. Bolton[1†], H. Daido[9], Y. Kato[10], K. Kondo[1], S. V. Bulanov[1,6,11], and M. Kando[1]

In various media the elementary components can emit traveling waves such as electromagnetic, gravitational or acoustic types. If these elementary emitters are synchronized, the resulting emission is coherent[1]. Moreover, the faster the emitters approach an observer, the more intense and directional their apparent emission is, with associated frequency increase[2]. Multi-stream flows[3] ubiquitously occur in media (such as with shock waves[4,5] and jets[6] in astrophysical and laboratory plasmas[4,7]) and produce fast moving density singularities[8,9], where high concentration and synchronism, in principle, can bring constructive interference[10]. However, a singularity emitting such characteristic coherent radiation has not been demonstrated yet. We show this general phenomenon in laser-driven relativistic plasma, which is an ideal medium for realizing these effects in the laboratory under controllable conditions. Our experiments and simulations reveal bright

---

[1] Kansai Photon Science Institute, National Institutes for Quantum and Radiological Science and Technology, 8-1-7 Umemidai, Kizugawa-city, Kyoto 619-0215, Japan. [2]Graduate School of Engineering, Osaka University, 2-1 Yamadaoka, Suita, Osaka 565-0871, Japan. [3]Joint Institute for High Temperatures of the Russian Academy of Sciences, Izhorskaja Street 13/19, Moscow 127412, Russia. [4]Institute for Academic Initiatives, Osaka University, Suita, Osaka, 565-0871, Japan. [5]P. N. Lebedev Physical Institute of the Russian Academy of Sciences, Leninsky Prospekt 53, Moscow 119991, Russia. [6]Moscow Institute of Physics and Technology (State University), Institutskii pereulok 9, Dolgoprudnyi, Moscow Region 141700, Russia. [7]Central Laser Facility, Rutherford Appleton Laboratory, STFC, Chilton, Didcot, Oxon OX11 0QX, UK. [8]Department of Physics, SUPA, University of Strathclyde, Glasgow G4 0NG, UK. [9]Applied Laser Technology Institute, Tsuruga Head Office, Japan Atomic Energy Agency, Kizaki, Tsuruga, Fukui 914-8585, Japan. [10]The Graduate School for the Creation of New Photonics Industries, 1955-1 Kurematsu-cho, Nishiku, Hamamatsu, Shizuoka 431-1202, Japan. [11]A. M. Prokhorov Institute of General Physics of the Russian Academy of Sciences, Vavilov Street 38, Moscow 119991, Russia.
[†]Present address: Chair of Experimental Physics and Medical Physics, Faculty of Physics, Ludwig-Maximilians-Universität München, Am Coulombwall 1, D-85748 Garching b. München, Germany

**coherent soft x-ray radiation from nanoscale electron density singularities in multi-stream plasma. They constitute a new compact x-ray source of ultrashort duration, demanded in numerous applications. In general, singularities can be bright sources of other types of traveling waves. Thus our findings open new opportunities in different fields of science. For example, gravitational wave[11,12] generation, as proposed in ultrahigh-energy accelerators[13,14], can be significantly enhanced by intentionally induced density singularities in the particle bunches. Further, we anticipate that multi-stream flows in cosmic media can produce intense bursts of coherent electromagnetic and/or gravitational waves, especially at longer wavelengths which facilitate constructive interference. We can then expect to observe more directional short wavelength bursts from cosmic emitters approaching at relativistic speeds. Thus, we present a new framework for interpreting a broad range of experimental results[15,16].**

In a physical medium, converging flows can lead to catastrophic changes of the medium characteristics, e.g. formation of structurally stable singularities such as long-lived density spikes, which are robust with respect to perturbations. Their existence, universality and structure are explained by catastrophe theory[8,9]. If the medium emits traveling waves, the singularity formation can cause a burst intensification of coherent emission because a high concentration of elementary emitters facilitates constructive interference of the emitted waves, Figure 1a. The coherent emission originates from regions with the size well below the emitted wavelength. Here the emitted wave phase should be a continuous function of the emitter coordinates and/or momenta. This condition is satisfied, e.g., when the elementary emitters are driven (excited) by an external wave. The resulting emission intensity scales quadratically[1] with the number of elementary emitters in the singularity, in contrast to the linear scaling of the intensity of incoherent emission from the background. If the singularity moves with a relativistic speed corresponding to the Lorentz factor $\gamma$, its emission, confined in a narrow angle of the order of $1/\gamma$ in the motion direction, becomes at least $\gamma^4$ times more intense[1,17]. It is essential that the singularity consists of elementary emitters continuously flowing through it. In this respect it fundamentally differs from the case of a compact bunch[18], which consists of the same elementary emitters. In both cases only the waves emitted from the location of the highest density interfere constructively. However, in contrast to a particle bunch, the singularity is determined by the multi-stream flow and has a non-local nature.

The concept formulated above can be called Burst Intensification by Singularity Emitting Radiation (BISER). Laser plasma is eminently suitable for its realization and study. An ultra-intense laser quickly ionizes matter creating relativistic plasma[19], where multi-stream flows and singularities emerge, e.g., at the longitudinal[20] and transverse breaking[21] of nonlinear Langmuir waves[22,23]. When a tightly focused laser pulse propagates in low-density plasma, Figure 1b, it pushes electrons forming a cavity in the electron density and a bow wave[24], Figure 1c. The resulting transverse displacement of electrons, Figure 1d, produces density singularities, Figure 1e, seen as a density spike and two sharp boundaries of the cavity wall and bow wave. Due to the modulations imposed by the laser with the scale of the laser wavelength, Figure 1c, the density spike oscillates emitting high-frequency electromagnetic radiation[10] in a cone around its velocity vector, analogously to a relativistic oscillating electric charge. Remarkably, the density singularities are robust with respect to laser-imposed modulation, because they correspond to structurally stable catastrophes[8,9]. Sharp outlines of the cavity wall and bow wave, Figure 1c, correspond to fold catastrophes. The density spike corresponds to a higher order cusp catastrophe at the joint of two folds. The cusp is located on a line encircling the laser pulse. A linearly polarized laser pulse pushes electrons stronger along the polarization axis, creating higher concentration and stronger modulation of electrons at the two opposite points on the cusp line. Since the radiation intensity at constructive interference quadratically depends on the electron number, these two points represent the strongest emitters.

We demonstrated the BISER concept in the interaction between a multi-terawatt femtosecond laser, Extended Data Figure 1, with the estimated peak irradiance ranging from $I_0 \approx 6 \times 10^{19}$ to $2 \times 10^{20}$ W/cm$^2$ and a supersonic helium jet with the electron density from $n_e \approx 1.4 \times 10^{19}$ to $6 \times 10^{19}$ cm$^{-3}$, using high spatial resolution soft x-ray diagnostics, Extended Data Figure 2-5. Under a wide range of experimental conditions, we observed coherent soft x-ray point-like emitters with brightness exceeding that of the background laser plasma by many orders of magnitude. The brightest emission corresponded to two emitters seen as bright spots in Figure 1b, Figure 2, and Extended Data Figure 2e, in the plane perpendicular to the observation direction. Their apparent size was below 1 µm, while the distance between them was from 10 to 20 µm. Prominent fringes in Figure 2 and Extended Data Figure 3b suggested that the transverse source size was not greater than 100 nm. For comparison, the initial laser spot size was about 10 µm, Extended Data Figure 1a, while an incoherent plasma emission was observed from a 0.5 mm region, Extended Data Figure 2c. The energy emitted into the acceptance angle and observable spectral range from 60 to 100 eV was up to ~100 nJ, corresponding to ~$7 \times 10^{11}$ photons. The typical divergence was ~20-30°. The spectra comprised high-order harmonics of optical frequency, Figure 3a,b, although sometimes harmonics were unresolvable. The radiation was spatially and temporally coherent, as evidenced by the spatial and spectral fringes, Figure 2 and Figure 3b, respectively.

We performed a particle-in-cell (PIC) simulation of an intense laser pulse propagating in plasma with parameters close to that of the experiment, Figure 4, Extended Data Figure 6-10, and Supplementary Movies M1-M3. A multi-stream electron flow created density spikes (seen in Figure 1c, which is a close-up of one of the density spikes in the middle panel in Figure 4). Two density spikes produced soft x-ray emission in the form of trains of short pulses (Figure 3c,d and Figure 4, bottom row), carrying a total energy of 120 nJ within the photon energy range from 60 to 90 eV propagating into the acceptance angle of the detectors. The individual short pulses corresponded to the density spike oscillations. The duration of these pulses, 170 attoseconds, was comparable with the bandwidth-limited 150 attosecond pulse estimated from the experimental spectrum within the same photon energy range, Figure 3d. The density spike's transverse size was 14 nm, full width at half maximum (FWHM).

An electromagnetic field near the electron bunch injected[25] into the cavity (Figure 4, right) mainly was not emitted, i.e. not converted into a traveling wave, because it was the field associated with the electric charge and current of the bunch[26]. Only the density spikes could produce high harmonics of the optical base frequency, because they emitted trains of short pulses, Figure 3c.

The sub-micron size of the observed emitters makes them unique among x-ray sources. Other sources, such as coherent and incoherent x-rays due to atomic transitions[27], recollision process[28], and nonlinear Thomson scattering[29,30], originate from regions with typical size from several to a few hundred micrometres, as determined by the laser focal spot and/or plasma length, Extended Data Figure 1.

Our findings reveal a high-power bright compact coherent x-ray source. Assuming a 170 as x-ray pulse duration (shown by the simulation) and restricting the pulse energy to 20 nJ within a single pulse of the train, we estimate the emitter pulse power as 0.1 GW. The estimated peak brightness at the wavelength of 18 nm is $10^{27}$ photons/mm$^2$·mrad$^2$·s in 10% bandwidth.

The BISER concept, on the one hand paves the way towards sources of traveling waves based on singularities formed in media under controllable conditions, on the other hand allows the revealing of singularities in multi-stream flows. Both of these aims are achieved in our experiments and numerical simulations. Remarkably, the source size is orders of magnitude smaller than the driving agent. In our case, it was smaller than the laser pulse and even its smallest characteristic scale (wavelength). In general, the apparent source size and divergence of coherent emission depend on the spectral range which is searched. A smaller source is seen at shorter wavelengths. From a broader perspective, the BISER concept creates a new framework for interpreting astrophysical observations, where media emitting electromagnetic and/or gravitational waves exhibit multi-stream flows and density singularities moving with relativistic speed, which can be the progenitors of gamma-ray bursts[15] and fast radio bursts[16]. The BISER concept can enrich the range of theoretically predicted waveforms used for the signal search[12] in the modern gravitational wave astronomy.

**Supplementary Information** is linked to the online version of the paper at www.nature.com/nature.

**Acknowledgments**: We thank the J-KAREN laser operation group. We acknowledge the financial support from JSPS Kakenhi JP 25287103, JP 25390135 and JP 26707031.

**Author Contributions**: A.S.P. and T.Zh.E. contributed to the research equally. The experiments were planned and prepared by A.S.P., M.K., H.D., K.O., E.N.R., D.N., T.I., M.N., N.H., T.K., P.R.B., K.K., led by A.S.P. and M.K., and performed by T.A.P., A.Ya.F., K.O., Y.H., H.Kotaki, D.N., H.Kiriyama, Y.F., and A.S. The experimental data were analysed by A.S.P., T.A.P., and A.Ya.F. and interpreted by A.S.P., T.Zh.E., T.A.P., A.Ya.F., H.D., J.K.K., T.K., P.R.B., K.K., Y.K., and S.V.B. T.Zh.E. performed the simulations and interpreted their results. A.S.P., T.Zh.E., T.A.P., and A.Ya.F. led the manuscript writing. All authors contributed to the final version of the manuscript.

**Author Information**: The data are available from the corresponding authors on reasonable request. Reprints and permissions information is available at www.nature.com/reprints. The authors declare no competing financial interests. Correspondence and requests for materials should be addressed to pirozhkov.alexander@qst.go.jp and timur.esirkepov@qst.go.jp.

## Methods

**Relativistic plasma.** Electron dynamics is relativistic when the electron velocity, $v$, becomes close to the speed of light in vacuum, $c$, so that the electron's Lorentz factor is substantially greater than unity, $\gamma = (1-v^2/c^2)^{-1/2} \gg 1$. The electron dynamics in the laser field becomes relativistic when the laser dimensionless amplitude is comparable to or greater than unity[19]. The dimensionless amplitude is defined as $a_0 = eE_0/m_e c\omega_0 = (I_0/I_R)^{1/2}$, where $I_R = \pi c^5 m_e^2/2e^2\lambda_0^2 \approx 1.37\times10^{18}$ W/cm$^2 \times (\lambda_0[\mu m])^{-2}$ for a linearly polarized field. The dimensionless amplitude characterizes the importance of relativistic effects in laser-plasma interaction. A laser is said to be relativistically strong when $a_0 \geq 1$. Here $E_0$, $I_0$, $\lambda_0$ and $\omega_0$ are the laser electric field, irradiance, wavelength and angular frequency, respectively; $e$ and $m_e$ are the electron charge and mass, respectively; $\lambda_0[\mu m]$ denotes the wavelength in micrometres. Note that in the caption of Figure 4 the value of $I_R = 2\times10^{18}$ W/cm$^2$ is specified for the laser wavelength of $\lambda_0 = 810$ nm.

The plasma density is characterized by its initial electron density, $n_e$, with respect to critical density, $n_{cr} = m_e\omega_0^2/4\pi e^2 \approx 1.1\times10^{21}$ cm$^{-3}/(\lambda_0[\mu m])^2$. If $n_e \ll n_{cr}$, the plasma is transparent to electromagnetic radiation (in the small amplitude limit)[1]; then the plasma is said to be underdense.

A laser beam undergoes relativistic self-focusing[31,32,33] when its peak power, $P_0$, exceeds the threshold of $P_{SF0} = P_c (n_{cr}/n_e)$, where $P_c = 2m_e^2 c^5/e^2 \approx 0.017$ TW, Ref.[33]. In the stationary self-focusing regime[34], the self-focused spot diameter and dimensionless amplitude can be estimated as $d_0 = \lambda_0(a_0 n_{cr}/n_e)^{1/2}/\pi$ and $a_0 = (8\pi P_0 n_e/P_c n_{cr})^{1/3}$, respectively.

A relativistically strong laser pulse propagating in underdense plasma excites wake waves exemplified by a longitudinal Langmuir wave[22].

A relativistically strong laser pulse produces a prominent bow wave when its focal spot is narrower than the threshold, $d < d_{BW} = 2\lambda_0(a_0 n_{cr}/n_e)^{1/2}/\pi$, Ref.[24]. This can be achieved by a tight focusing of the laser beam or via relativistic self-focusing.

**Laser.** We used the multi-terawatt femtosecond J-KAREN laser[35] with the central wavelength of $\lambda_0 \approx 810$ nm. The laser pulses were linearly polarized. The pulse energy was controllably varied from $\mathcal{E}_L = 0.1$ to 0.7 J. The soft x-ray emission from point-like sources was observed with laser pulse energies higher than 0.18 J. The temporal pulse shape, Extended Data Figure 1a, was measured with the self-referenced spectral interferometry[36], resulting in the FWHM duration of about 30 fs and effective duration of $\tau_{Eff} \approx 35$ fs. Here $\tau_{Eff} = \int P_n(t)dt$ is the area under the normalized power curve $P_n(t)$. The peak power $P_0$ calculated as $P_0 = \mathcal{E}_L/\tau_{Eff}$ varied from ~3 to 20

TW. The pulses were focused with an f/9 off-axis parabola. The focal spot measured in vacuum, Extended Data Figure 1b, had the FWHM spot diameter $d_{VAC} \approx 10$ μm with the Strehl ratio of about 0.3. The estimated irradiance in the absence of plasma varied from $I_{Vac} \approx 2 \times 10^{18}$ W/cm$^2$ at the smallest pulse energy to $I_{Vac} \approx 10^{19}$ W/cm$^2$ at the highest one. Correspondingly, the dimensionless laser amplitude varied from $a_{0,Vac} \approx 0.9$ to 2.2. The laser parameters fluctuated slightly day to day; the actual values were used in the data analysis. The laser pulse and spot shape were good enough to prevent filamentation development, as confirmed by the plasma channel image, Extended Data Figure 1c.

It was essential that the laser beam underwent relativistic self-focusing, because the laser peak power, $P_0$, significantly exceeded the relativistic self-focusing threshold $P_{SF0}$. This effect greatly enhanced the laser irradiance and, correspondingly, the laser dimensionless amplitude. The achieved peak intensity was estimated to be in the range from $I_0 \approx 6 \times 10^{19}$ to $2 \times 10^{20}$ W/cm$^2$ for the parameters of the experiment. This effect also facilitated the tightest focusing which satisfied the condition for the bow wave formation by the laser pulse. The self-focused spot diameter, $d_0$, estimated according to the stationary self-focusing regime[34], was safely two times narrower than the bow wave formation threshold, $d_{BW}$.

**Gas jet**. We used a 1-mm orifice diameter supersonic helium gas jet with the Mach number of 3.3. The gas density distribution was measured for several backing pressures using optical interferometry with Ar gas. The Ar gas had negligible clusterization under our conditions (Hagena's parameter[37] values $\Gamma^* < 7 \times 10^4$), and thus its atomic density was nearly identical to He with the accuracy of several percent[38]. The estimated peak density error was ≈ 30% determined by the interferometry noise and reconstruction process.

The peak electron density in the helium jet, controlled by the backing pressure, was varied from $n_e \approx 1.4 \times 10^{19}$ to $6 \times 10^{19}$ cm$^{-3}$. The density was calculated as twice the atomic density because helium is fully ionized for the intensities used in the experiment[39]. This density was much lower than the critical density $n_{cr} \approx 1.7 \times 10^{21}$ cm$^{-3}$ for our laser's central wavelength of $\lambda_0 \approx 810$ nm.

**Diagnostics setup.** We sequentially used two detectors, the Spectrograph and the Imager, placed behind the gas jet at the angle of 13° with respect to the laser propagation. Both detectors used the same near-normal-incidence imaging mirror (schematically represented by "imaging optics" in Figure 1b). The mirror had a 48 mm open diameter and a 500 mm radius of curvature. Due to the near-normal incidence, the mirror provided a large acceptance angle, collecting radiation propagating in the ±5° cone around the angle of 13° with respect to the laser axis in the polarization plane (in other words, the cone is bounded by the angles from 8° to 18° in that plane). The mirror formed low-aberration magnified images of the source on the imaging sensors, either a back-illuminated Charge-Coupled Device (CCD) for the Spectrograph, Extended Data Figure 2, or a LiF crystal for the Imager, Extended Data Figure 3. As no single material can afford efficient normal-incidence reflection in the soft x-ray spectral region, we used an aperiodic Mo/Si multilayer coating[40] designed for approximately uniform 11% reflectivity in the 12.4 to 20 nm spectral region[41], corresponding to the photon energy range of 60 to 100 eV. To avoid detector exposure from laser and out-of-band plasma radiation, we used two free-standing optical blocking filters, specifically two 0.2 μm thick multilayer Zr/Al filters[42] in the case of the Spectrograph and 0.1 μm Zr and 0.2 μm Zr/Al filters in the case of the Imager. A 0.4 T, 5 cm long magnet deflected charged particles away from the sensors.

**Spectrograph**. This detector, Extended Data Figure 2, registered spatially-resolved spectra and spectrally-integrated images of electromagnetic radiation in the soft x-ray spectral region with the photon energies from 60 to 100 eV (corresponding to the wavelengths from 12.4 to 20 nm). The spectrograph throughput is shown in Extended Data Figure 4a. The sharp edge at 12.4 nm was due to the Silicon L absorption edge, while the roll-off at $\lambda > 20$ nm was due to the reduced filter transmission. A spectral resolution of about 0.1 nm was attained without a slit due to the small source size. The spatial resolution reached several to 10 μm in the object space. Both spectral and spatial resolutions were determined by the 13.5 μm CCD pixel size and geometrical aberrations.

**Imager**. This detector, Extended Data Figure 3, recorded spectrally-integrated images of soft x-ray plasma emission covering the same spectral region as the Spectrograph, from 60 to 100 eV, Extended Data Figure 4b. The Imager provided much higher, better than 1 μm, spatial resolution because we used smaller incidence angle and a LiF crystal imaging sensor instead of the CCD. The high-frequency radiation caused formation of colour centres in the LiF crystal, recording the source images. These images were read out after the experiment with a fluorescence microscope[43]. Being excited by UV radiation the colour centres emitted visible light recorded by the microscope. The estimated depth of focus for the Imager was smaller than 10 μm, in the case of <1 μm objects. Therefore the observation of two point-like sources indicated that the distance between them in the direction of observation was well below 10 μm; sources with larger longitudinal separation would appear blurred (defocused).

**Experimental data.** Using the Spectrograph's ability of every laser shot readout, we performed a survey of the laser-plasma interaction varying the laser pulse energy, gas jet pressure, and position of the laser focus inside the gas jet. Over a broad range of experimental conditions stated in the main text, we observed bright soft x-ray radiation, always emitted from point-like sources, as shown in Extended Data Figure 2. The harmonics were not always resolved, which might be connected either with a single emission spike in the time domain[44] or defocusing, because even a few-tens μm shift of the source position resulted in the spectral image blurring and corresponding lower spectral resolution.

The angular distribution of the soft x-ray emission was measured using the Spectrograph in the strongly defocused mode, when the pixel position on the CCD corresponded to the observation angle rather than position in the object space. The precise angular scale calibration was performed using the shadow of the 150 μm period parallel support mesh of the transmission grating, Extended Data Figure 2b. In the laser polarization plane, the off-axis radiation typically extended to 10-13° and eventually exceeded 18°, our detectors' limit. The full divergence estimated from these data was 20-30°. In the perpendicular plane, the typical divergence was 10°.

The pulse energy and absolute number of photons in the soft x-ray spectra were estimated using the calculated mirror reflectivity which agreed with measurements at several wavelengths[41], measured filters transmission[42], calculated transmission grating efficiency[45,46], and CCD efficiency and gain provided by the manufacturer. In this way the Spectrograph throughput, Extended Data Figure 4a, was obtained. Extended Data Figure 5 gives the spectrum in absolute units of the stronger emitter of the shot shown in Figure 3a and Extended Data Figure 2.

As the CCD pixel size in the Spectrograph did not allow achieving sufficient spatial resolution, we set experimental parameters corresponding to the strongest soft x-ray emission and used the Imager for obtaining the high-resolution images of the emitters, Figure 2. The stronger emitter is shown in Extended Data Figure 3b. The determination of its size, apparently 1.5 μm by 0.8 μm, was limited by the imaging system imperfections. However, the prominent fringes seen in the lineout, Extended Data Figure 3b, suggested that the transverse size was substantially smaller. Indeed, physical optics propagation modelling provided a conservative estimate of the transverse source size to be not greater than 100 nm; for larger sizes the modelling showed much fainter fringes.

**Simulation.** The simulation was done with the multi-parametric multi-dimensional code REMP, based on the Particle-in-Cell method[47] and the density decomposition scheme[48]. The simulation configuration was two-dimensional. The simulation box size was $104\lambda_0$ and $180\lambda_0$ in the longitudinal and the transverse direction, respectively. Along those directions the mesh size was, respectively, $\lambda_0/400$ and $\lambda_0/96$; the time step was $2.426\times10^{-3}\ \lambda_0/c$. The simulation was performed using the moving window technique[49]. In this technique, the simulation box (window) moves with the speed of light, $c$, in the longitudinal direction, in order to observe the laser pulse evolution and corresponding phenomena, neglecting processes far behind the laser pulse. The processes left behind the window boundaries could not influence the dynamics inside the window, because their influence propagated with the velocity not greater than the speed of light, $c$. The plasma was modelled by electrons on a background of immobile ions. In this approximation, the ion dynamics was neglected. Owing to large ion inertia, the ion response to the laser field and plasma wake fields was negligible on the timescale of the moving window. The maximum number of quasi-particles representing electrons was $2.2 \times 10^9$.

The laser pulse had a Gaussian profile with FWHM duration of 38 fs and FWHM spot size of 13 µm with respect to irradiance. The laser pulse was linearly polarized in the transverse direction, in the plane of the simulation box. The shape of the laser pulse in terms of dimensionless amplitude was

$$A_0(t,x,y) = a_0 \left(\frac{1}{2}\right)^{(2(x-t)/L_x)^2} \left(\frac{0.249}{0.25+(y/L_y)^2} + 4 \times 10^{-3} e^{-2.77(y/L_y)^2}\right),$$

which was an approximation of the result of the laser pulse measurements. Here $L_x=20\lambda_0$, $L_y=25\lambda_0$.

The plasma density profile was approximation fitted from the gas jet measurements. In the transverse direction the plasma density was almost flat, while in the longitudinal direction it was given by the formula

$$n_e(x) = n_{e,\max}\left(\frac{0.357}{1 + 3.41\,(x/S_x)^2} + 0.643 e^{-2.95(x/S_x)^2}\right),$$

where $S_x = 525\lambda_0$.

Extended Data Figure 6 shows the laser pulse evolution (top panels, thick curves) in plasma (top panels, blue colour scale for electron density), and corresponding high-frequency emission going into the cone $\pm 18°$ around the laser axis (bottom panels, orange-red colour scale). See also the Extended Data Figure 7-10 and Supplementary Movies M1-M3. In M1, the high-frequency emission going only into the experimental acceptance angle is shown. In M2, M3, the aperture is wider, from $-18°$ to $18°$ around the laser axis. The movie M3 presents the laser propagation in the global window, in contrast to M1, M2 showing the processes in the moving window.

During the laser pulse propagation in plasma, the laser pulse waist decreased while its irradiance increased due to relativistic self-focusing. The laser pulse pushed electrons aside (away from the laser axis), creating an almost void cavity in the electron density and an outgoing bow wave (top rows in Figure 4 and Extended Data Figure 6). The electrostatic potential of the cavity pulled background electrons transversely toward the laser axis. This process is called transverse wave breaking (TWB)[21]. The interpenetrating transverse flows of these electrons formed a characteristic outline (folding), denoted in Extended Data Figure 6 as TWB. At a later time, longitudinal wave breaking[20] occurred, leading to the injection[25] of an electron bunch into the cavity (the rightmost panel of the top row in Extended Data Figure 6).

The electron density spike was formed at the location, where the bow wave detached from the cavity wall. The spike size was 14 nm FWHM, Extended Data Figure 7. The spike moved forward entering the higher amplitude regions of the laser pulse, where it started to emit high-frequency radiation in the form of short pulse trains. With the change of the density spike motion direction, its emission direction accordingly changed. An x-ray searchlight formed in this way illuminated the detector acceptance angle at a certain time. Both the electron spikes emitted high-frequency radiation in a relatively wide angle around the laser axis resulting in two point-like sources seen in the acceptance angle of the Imager and Spectrograph.

We note that the main portion of the high-frequency electromagnetic field around the electron bunch (in the rightmost panels of Extended Data Figure 6) was not emitted because it was the field associated with the electric charge and current of the bunch[26]. In addition, as seen from the simulation, only the density spikes could produce high harmonics of the optical base frequency, because they emit trains of short pulses, Figure 3c.

**Brightness estimate.** We estimate the brightness of singular soft x-ray source using the measured spectrum (Extended Data Figure 5), corresponding ~5° effective angular width measured with the Spectrograph in the intentionally defocused mode calibrated with the shadow of parallel support structure (Extended Data Figure 2b), and time evolution from the PIC simulation (Figure 3c). The source size in the laser polarization plane, 14 nm, was estimated from the simulation (Extended Data Figure 7), while in the perpendicular plane it was estimated from the experiment (Figure 2).

**Figures**

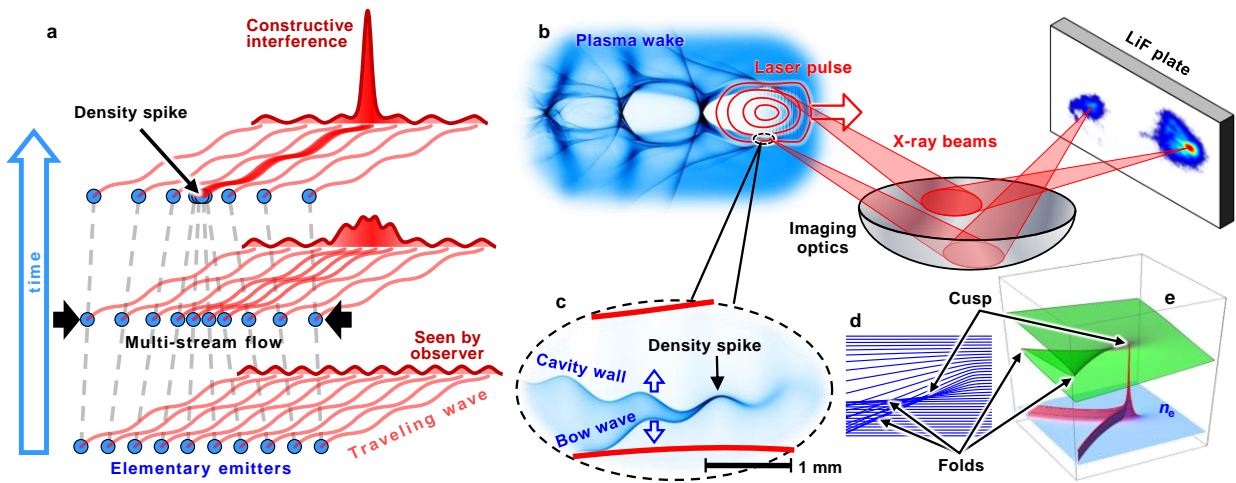

**Figure 1 | Formation of a singularity (density spike) and corresponding burst intensification of emitted radiation. a**, Multi-stream converging flow consisting of elementary emitters eventually forms density spike generating burst of coherent emission. **b**, In the experiment, the near-infrared laser pulse propagates in plasma creating density spikes, which produce x-rays collected by imaging optics and recorded by a LiF plate. **c**, A density spike appears in a multi-stream flow of electrons pushed aside by the laser field (PIC simulation). **d**, **e**, A multi-stream flow model, where the electron transverse displacement forms regions of high concentration, corresponding to density catastrophes – point-like cusp and folds.

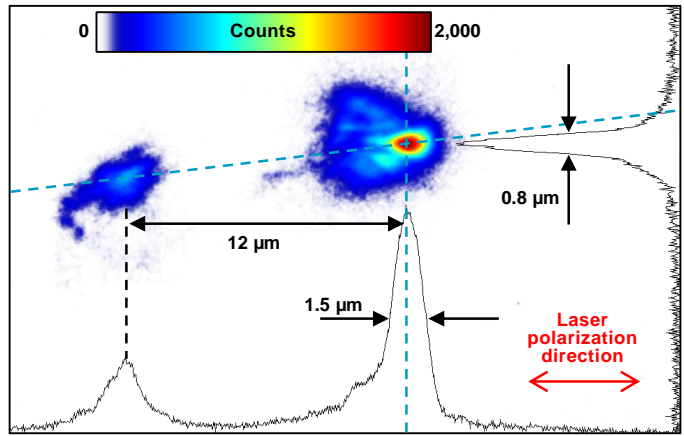

**Figure 2 | Singular emitters revealed in experiment.** Single-shot image produced on the LiF plate by photons with the energy from 60 to 100 eV. The solid lines show lineouts along the light dashed lines.

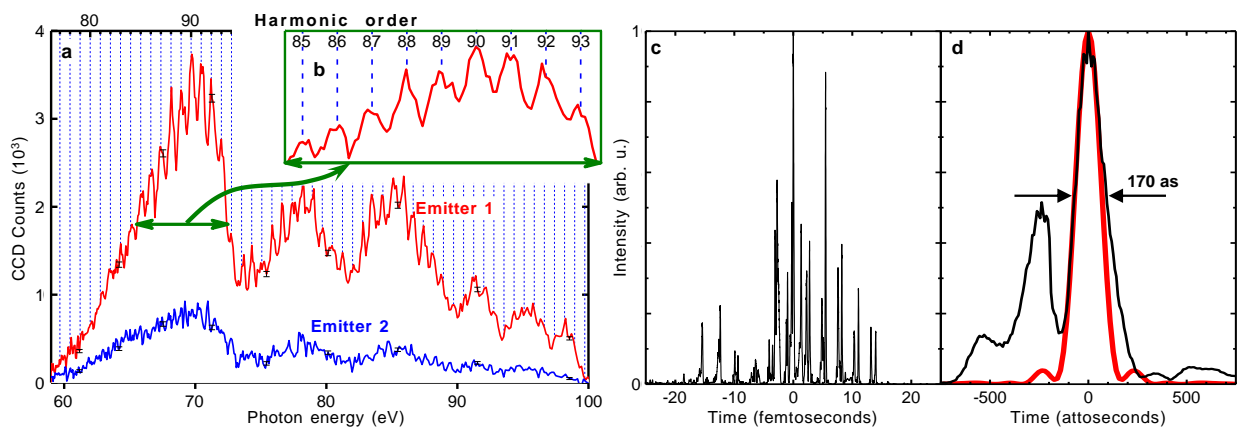

**Figure 3 | Properties of singular emitters. a**, Spectra of two point-like emitters obtained in a single shot (experiment). **b**, Inset shows harmonic structure in more detail. **c**, Temporal structure of the harmonic pulse exhibiting an attosecond pulse train (PIC simulation). **d**, The strongest attosecond pulse in the train (**c**) with the duration of 170 as. The red line shows the bandwidth-limited 150 attosecond pulse estimated from the experimental spectrum in the 60 to 90 eV range, panel **a**.

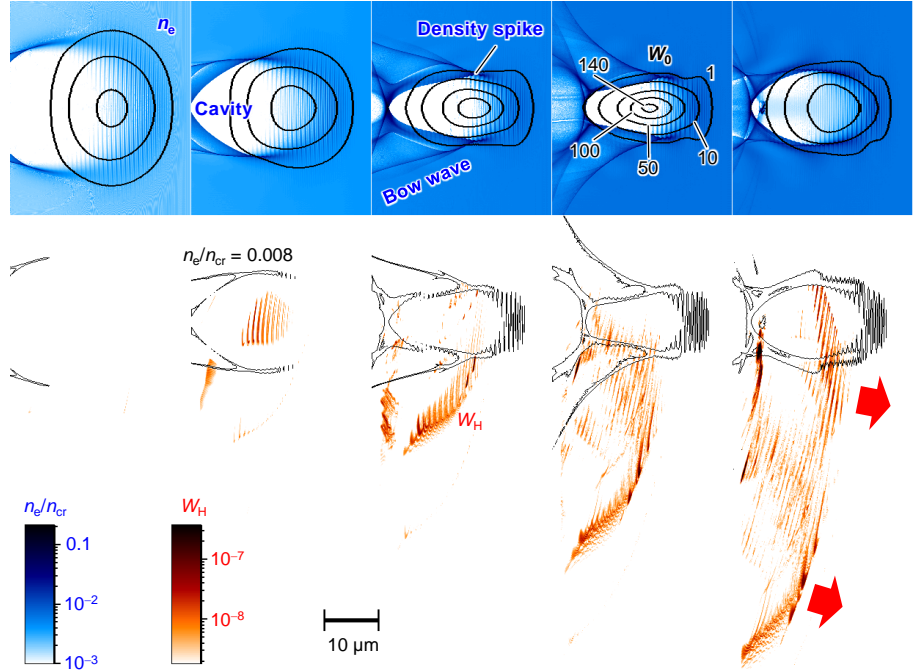

**Figure 4 | PIC simulation.** Top: the laser pulse, represented by the curves of irradiance $W_0$ in units of $I_R=2\times10^{18}$ W/cm$^2$, propagates in inhomogeneous plasma indicated by the electron density $n_e$ (blue, in units of $n_{cr}=1.7\times10^{21}$ cm$^{-3}$). Bottom: the irradiance $W_H$ of the high-frequency electromagnetic field with photon energy from 60 to 90 eV emitted between the angles of 8° and 18° (corresponding to the acceptance angle in the experiment) with superimposed electron density curve of $n_e/n_{cr}=0.008$. The time between panels is 270 fs. The rightmost panel corresponds to the centre of the jet.

# Extended Data

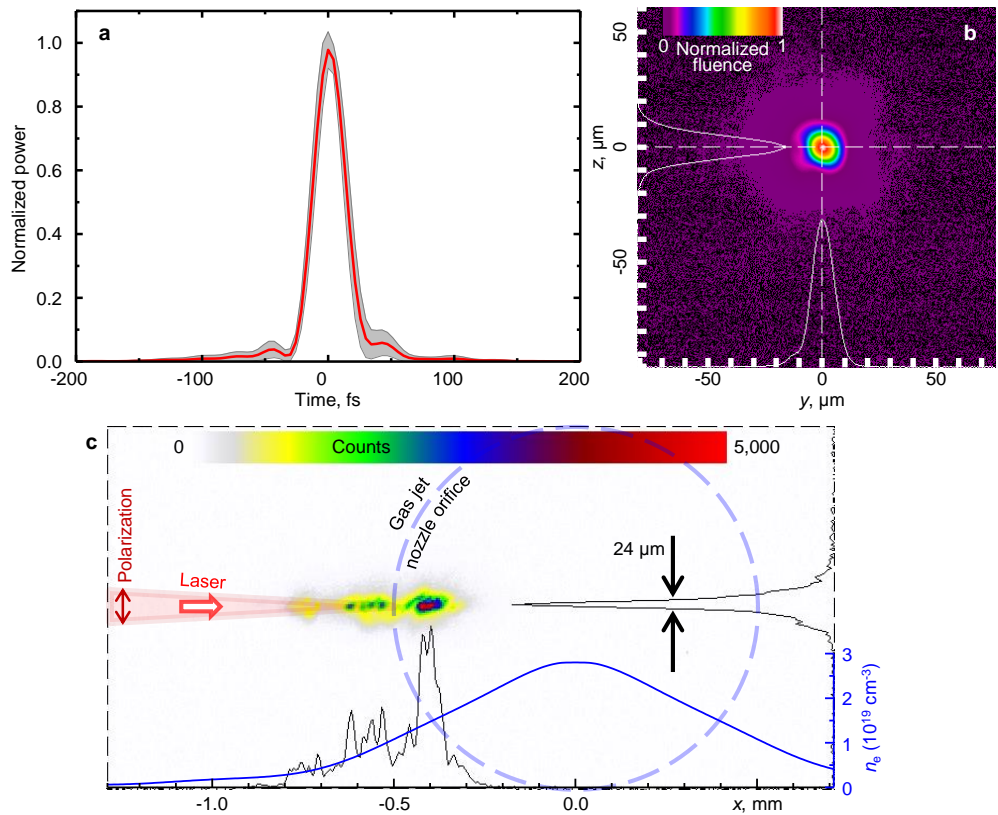

**Extended Data Figure 1 | Laser pulse properties. a**, Pulse shape measured with self-referenced spectral interferometry[36]; the thick red line shows the average, the grey area depicts the standard deviation over 113 single shots. **b**, Focal spot shape, the magnified image on a Charge Coupled Device (CCD) obtained with an achromatic doublet lens. **c**, The laser pulse irradiating the gas jet formed a plasma channel image (yellow-green-blue-red colour scale) which has no signs of filamentation. The dark-red arrow indicates the laser polarization direction.

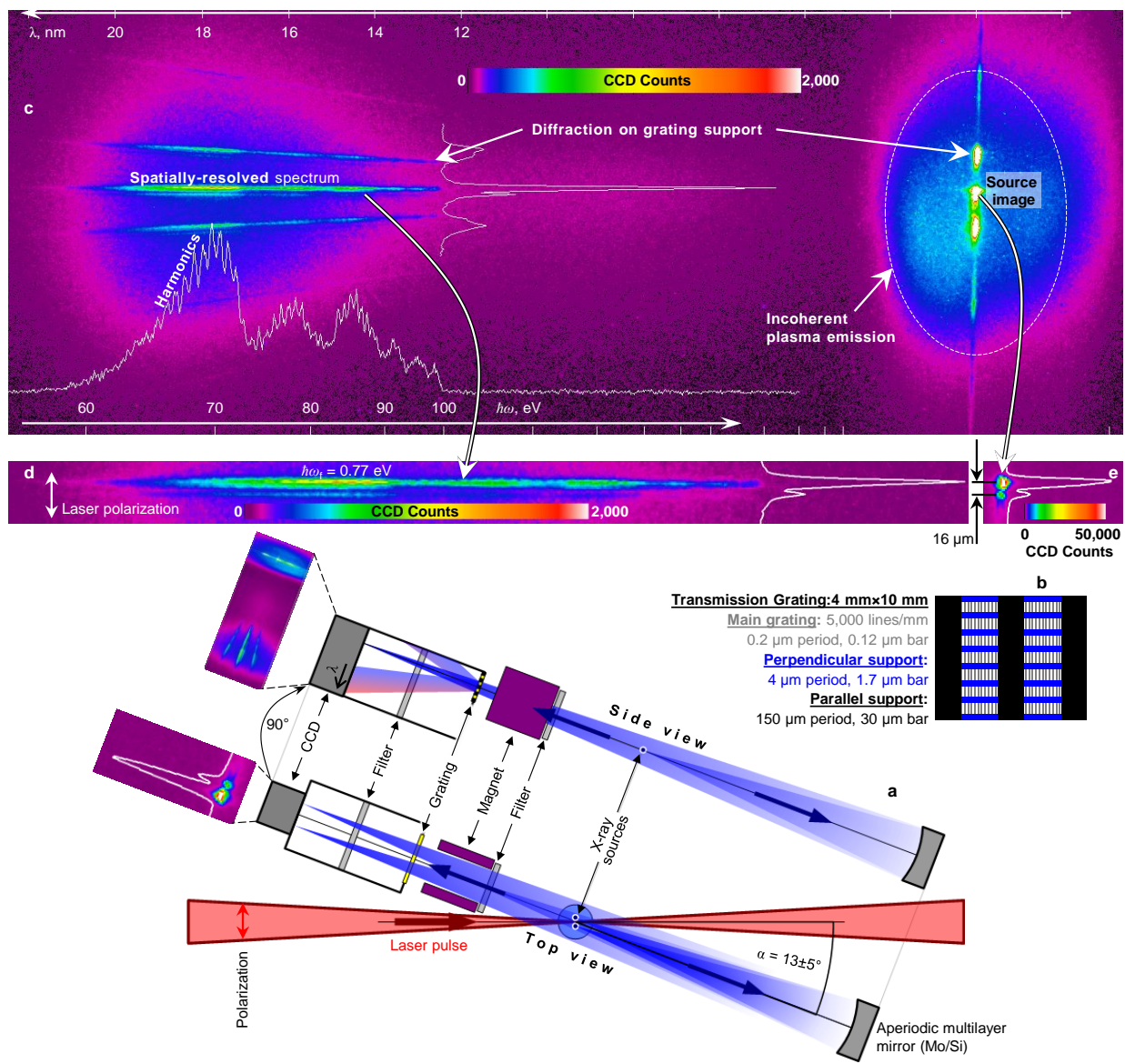

**Extended Data Figure 2 | Spectrograph and its data. a**, Experimental setup schematic of the Spectrograph (not to scale). The laser pulse irradiated the gas jet producing an x-ray emission. The red arrow denotes the laser polarization direction. The spherical mirror with aperiodic Mo/Si multilayer coating, set at the incidence angle of 1.8°, imaged the soft x-ray source to the back-illuminated Charge-Coupled Device (CCD) with the magnification of M = 6.85. Two Zr/Al optical blocking filters, 0.2 μm thick, rejected laser and out-of-band plasma

radiation. The magnet deflected charged particles. The transmission diffraction grating was used to obtain spectrally-resolved source images. In addition to the 5000 lines/mm main grating, there was also a 250 lines/mm periodic support mesh, **b**, which caused diffraction in the perpendicular direction, helping spectral identification. The parallel support mesh with the period of 150 μm was used for angular scale calibration in an intentionally defocused mode. An example of the raw data is shown in **c**; it corresponds to the shot shown in Figure 3a. The main grating formed the spectral decomposition in the horizontal direction, while the vertical direction represents the diffraction on the support mesh perpendicular to the main grating. Two sources of soft x-ray are seen in the first (horizontal) diffraction order, **d**, and in the zeroth spectral order, **e** (note the dynamic range difference between the panels **d** and **e**). The dashed ellipse in **c** denotes a dim incoherent plasma emission from area spanning about 500 μm. The spectrum lineout in **c**, corresponding to the highest signal in the first (horizontal) diffraction order, exhibits two kinds of oscillations: slow ones being determined by the spectrograph throughput, Extended Data Figure 4a, and higher-frequency ones corresponding to the high-order harmonics of the optical base frequency (seen also in Figure 3a,b).

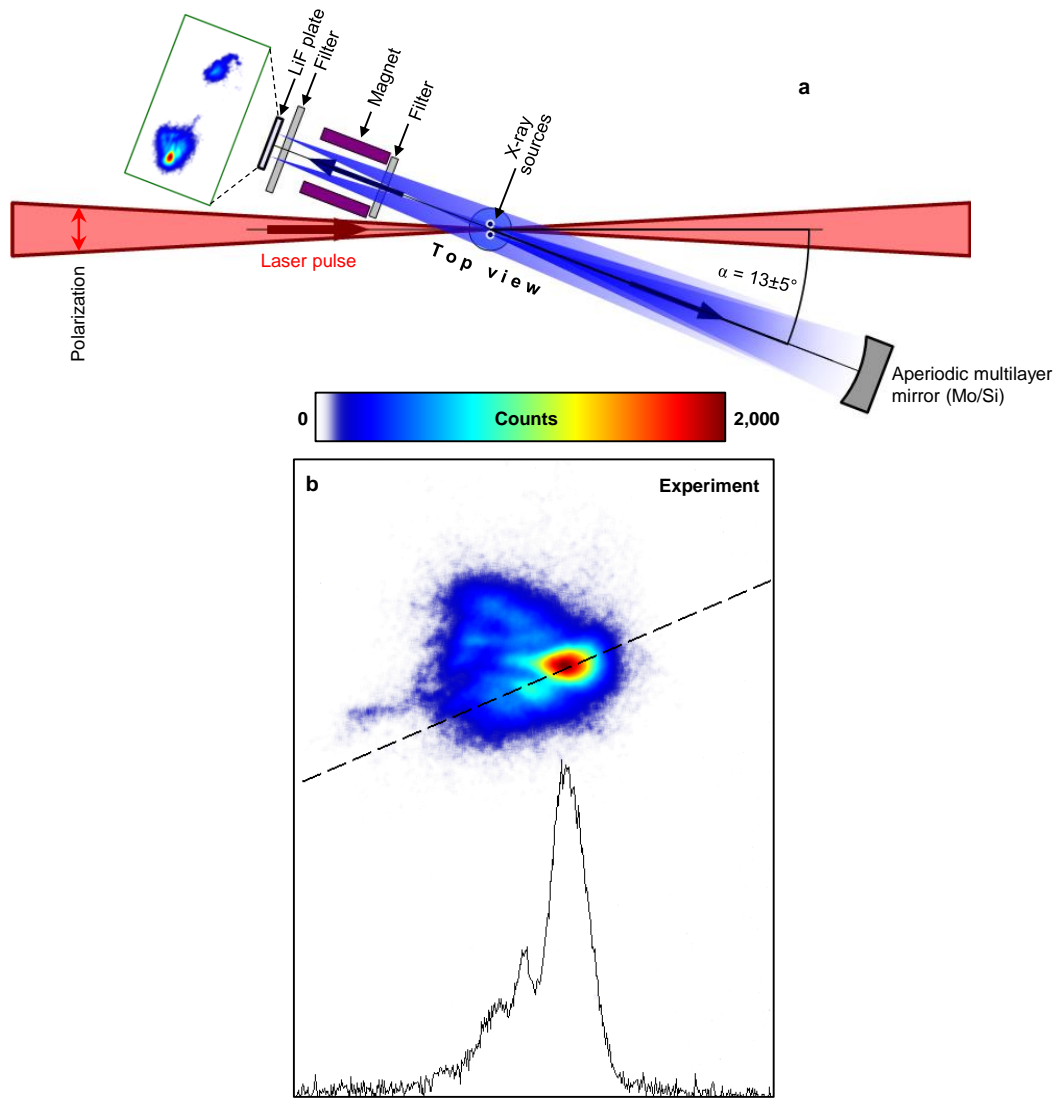

**Extended Data Figure 3 | Imager and its data.** Experimental setup schematic of the Imager (not to scale), **a**. The spherical mirror with aperiodic Mo/Si multilayer coating, set at the near-normal incidence angle of 1.04°, imaged the soft x-ray source onto a high-resolution LiF crystal imaging sensor with the magnification of M = 5.53. The optical blocking filters were 0.1 μm Zr and 0.2 μm Zr/Al. The example of the data is shown in Figure 2. Its stronger emitter image is shown in panel **b**; the lineout along the dashed line reveals fringes.

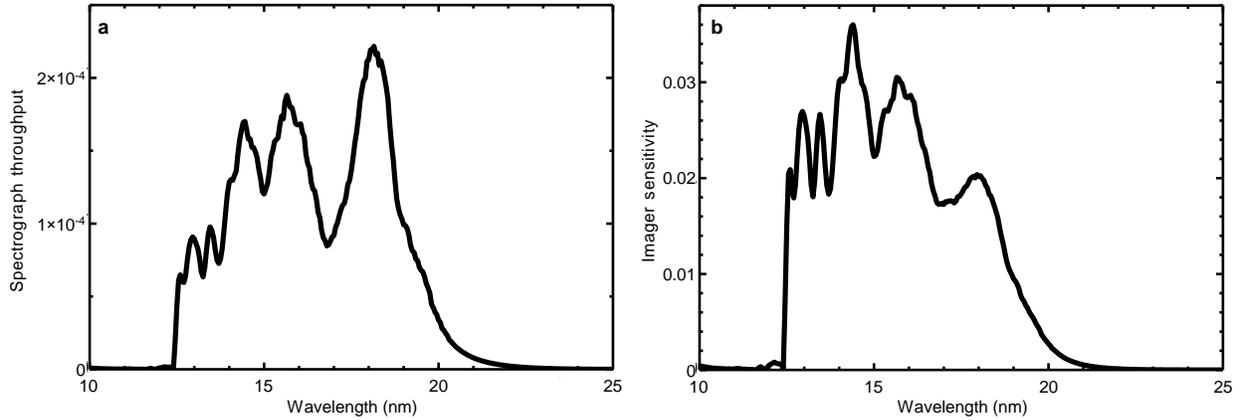

**Extended Data Figure 4 | Spectral ranges of the Spectrograph and Imager. a**, The Spectrograph throughput, i.e. the product of the Mo/Si multilayer mirror reflectivity, transmission of two Zr/Al filters, grating efficiency, and CCD quantum efficiency. **b**, The Imager spectral sensitivity curve, i.e. the product of the Mo/Si multilayer mirror reflectivity and transmission of Zr and Zr/Al filters.

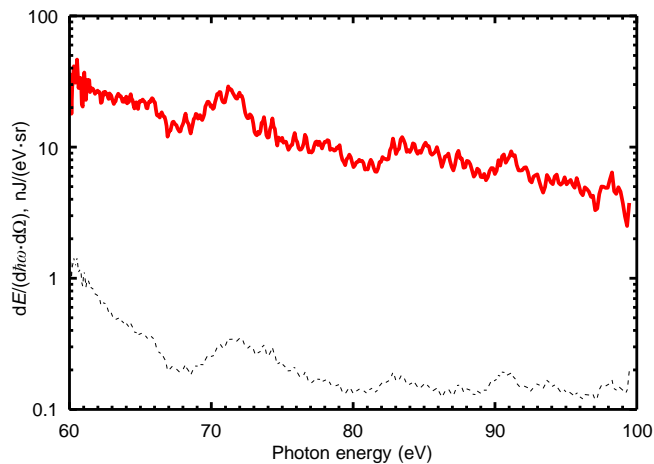

**Extended Data Figure 5 | Singular emitter spectrum in the absolute units.** The spectrum is obtained in the same shot as shown in Figure 3a and Extended Data Figure 2c-e. The dotted line represents the noise level.

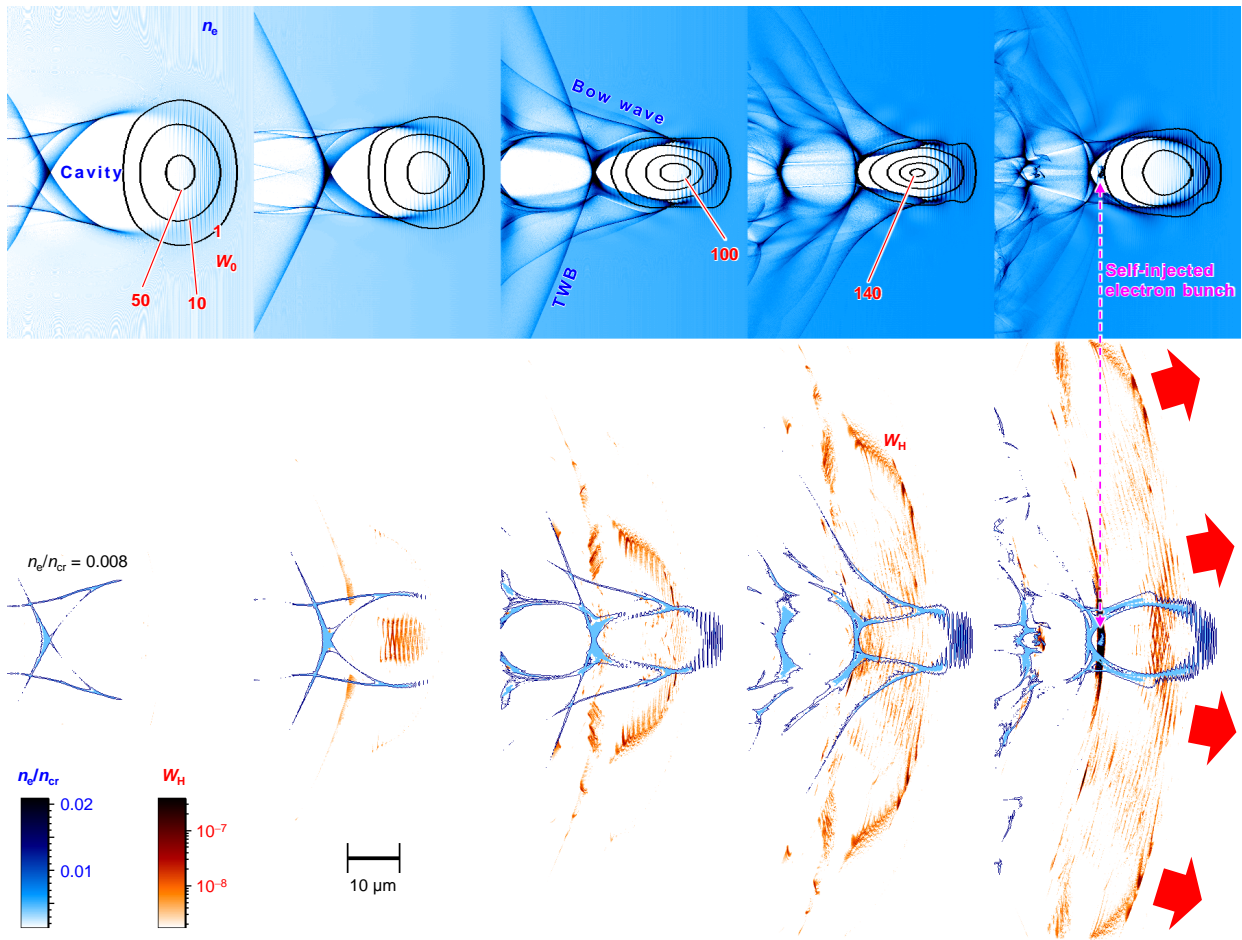

**Extended Data Figure 6 | PIC simulation: the laser pulse evolution and the BISER.** The panels show the same quantities and time moments as in Figure 4. Here the radiation propagating into a cone spanning from −18° to +18° is presented, in contrast to Figure 4, where only radiation propagating into the acceptance angle of the Imager and Spectrograph is shown, i.e. from 8° to 18° on one side of the laser axis. The scale of the electron density in the top row is linear rather than logarithmic as in Figure 4. TWB in the middle top panel stands for the Transverse Wave Breaking[21].

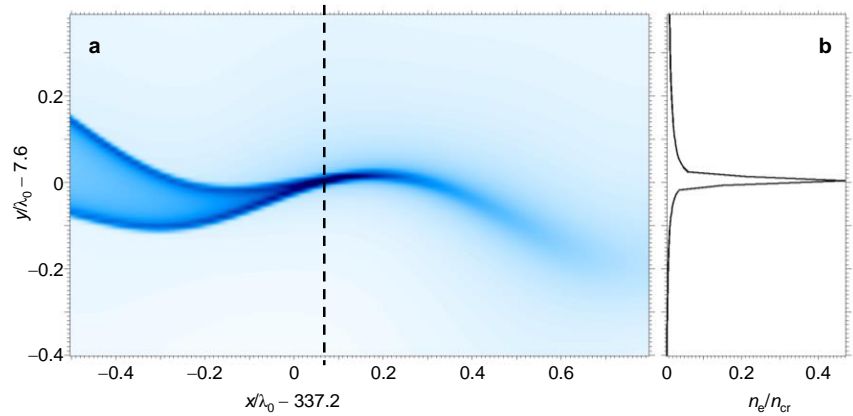

**Extended Data Figure 7 | PIC simulation: close-up of the electron density spike.** The region shown corresponds to the upper spike of the middle panel of Extended Data Figure 6.

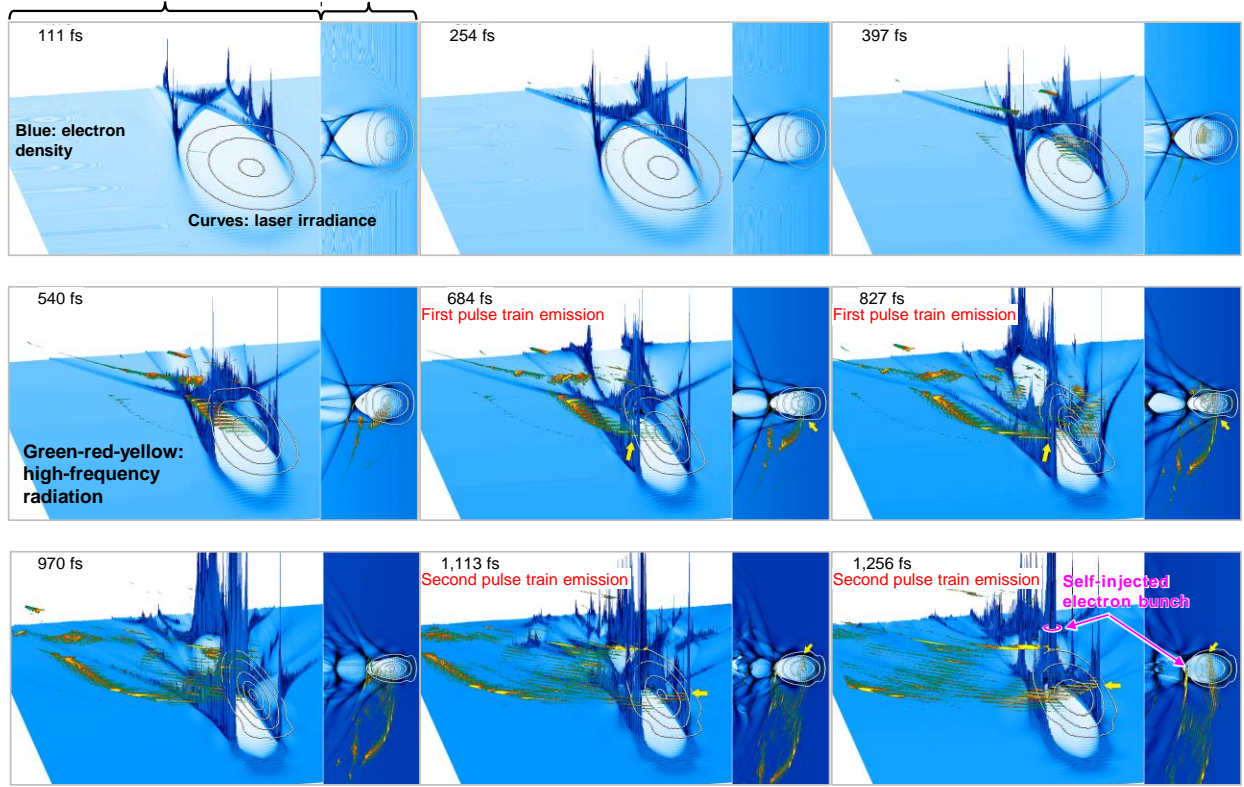

**Extended Data Figure 8 | Selected frames from Supplementary Movie M1.**

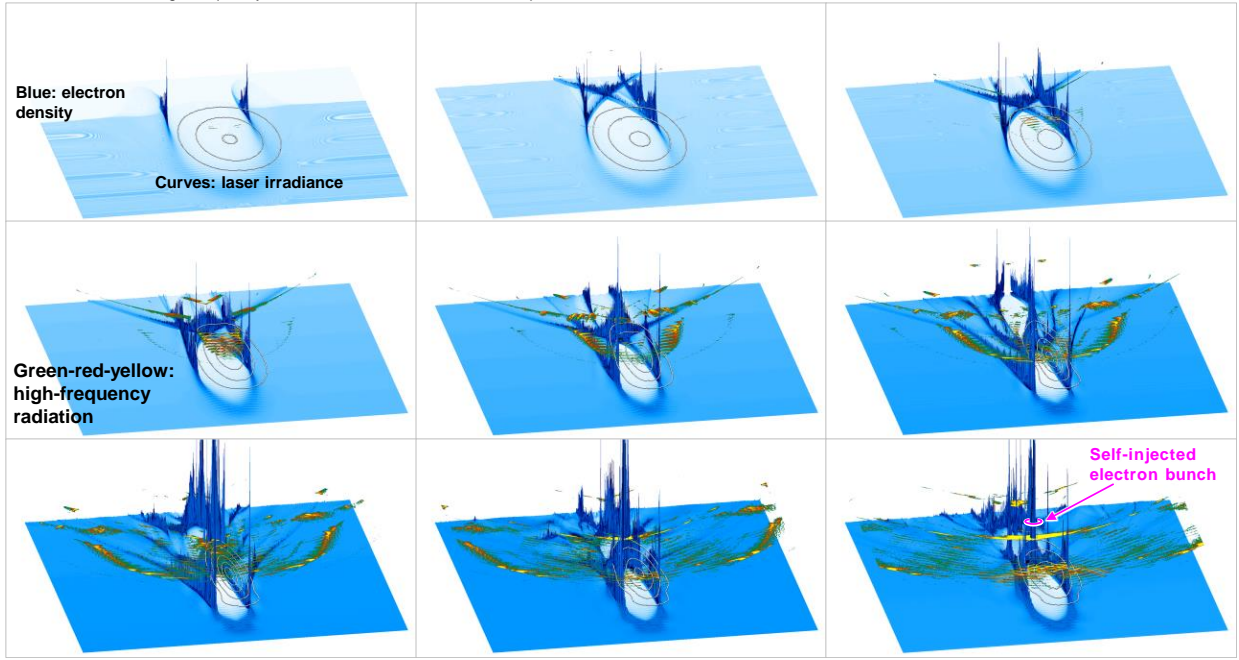

**Extended Data Figure 9** | Selected frames from Supplementary Movie M2.

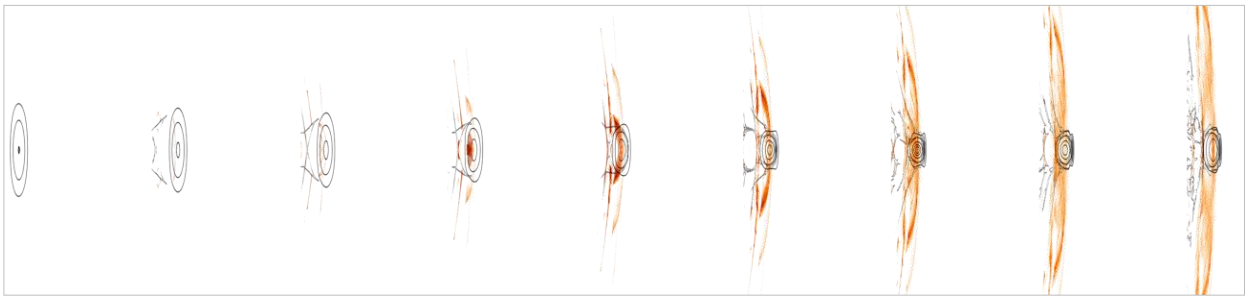

**Extended Data Figure 10** | Selected frames from Supplementary Movie M3.

# Supplementary Information

**Supplementary Movie M1. PIC simulation: the laser pulse evolution in plasma and the BISER going into the experimental acceptance angle.** High-frequency emission in the aperture bounded by the angles from 8° to 18° in the plane of laser polarization, in the window moving with the speed of light along the laser axis. The blue surface represents the electron density. Thick white curves show the laser irradiance levels of $W_0 = 1, 10, 50, 100, 140$, in units of $I_R = 2 \times 10^{18}$ W/cm$^2$, as in Figure 4. The green-red-yellow colour scale reveals the high-frequency emission with photon energy from 60 to 90 eV; the two strongest emission moments are denoted by the yellow arrows and signs "First pulse train emission" and "Second pulse train emission". Although the emission of each train continues for a few hundred femtoseconds, the full duration of the resulting attosecond pulse train is several femtoseconds, because the emitter itself moves with the relativistic velocity, i.e. almost catching up with the outgoing radiation. We also note that at the presented (down-sampled) resolution, the pulses of high-frequency emission are invisible until they sufficiently diverge. When their intensity distribution starts to occupy more than several pixels, they become visible. This explains why the apparent intensity of short pulses sometimes increases with time (larger area looks more intense than smaller one, especially near the saturation).

**Supplementary Movie M2. PIC simulation: the laser pulse evolution in plasma and the BISER going into angles from −18° to 18° near the axis**. Curves and colour scales are the same as in **Supplementary Movie M1**.

**Supplementary Movie M3. PIC simulation: the emission from the moving singularity**. Each frame corresponding to the moving window, obtained in the simulation, appears in its right place in the global window. In order to ease the observation, the aspect ratio is set to 1/3 = horizontal/vertical. Thin curves correspond to the electron density constant value of $n_e=0.008n_{cr}$, where $n_{cr}=1.7\times10^{21}$ cm$^{-3}$. Thick oval-like curves show the laser irradiance levels of $W_0 = 1, 10, 50, 100, 140$, in units of $I_R=2\times10^{18}$ W/cm$^2$. The orange-red colour scale shows the high-frequency emission with photon energy within the 60 to 90 eV spectral range propagating into the angles from −18° to 18° with respect to the laser axis in the plane of laser polarization.